\documentclass{article}

\usepackage{amssymb,amsfonts,amsmath}
\usepackage{cite,enumerate,float}
\usepackage{color}
\usepackage{tikz}
\usetikzlibrary{arrows,snakes,backgrounds}

\def\be{\begin{eqnarray}}
\def\ee{\end{eqnarray}}
\def\nn{\nonumber}

\def\p{\partial}
\def\tr{{\rm tr}\,}

\definecolor{red}{rgb}{1,0,0}
\definecolor{orange}{rgb}{1,0.5,0}
\definecolor{violet}{rgb}{0.7,0,1}

\hoffset= - 1.0in         % switch off for draft style
\begin{document}

\hfill MIPT/TH-21/22

\hfill ITEP/TH-27/22

\hfill IITP/TH-24/22

\bigskip

\centerline{\Large{
Equating Schur Functions
}}

\bigskip

\centerline{\bf A.Morozov }

\bigskip

\centerline{\it MIPT,  NRC ”Kurchatov Institute”, ITEP  \& IITP, Moscow, Russia }

\bigskip

\centerline{ABSTRACT}

\bigskip

{\footnotesize
We wonder if there is a way to make all Schur functions in all representations equal.
This is impossible for fixed value of time variables,
but can be achieved for averages.
It appears that the corresponding measure is just Gaussian in times,
which are all independent.
The generating function for the number of Young diagrams
does not straightforwardly appear as a product, 
but is reproduced in a non-trivial way.

}

\bigskip

\bigskip

{\bf 1.} Schur functions and their close relatives are acquiring a sort of a central position 
in advanced theoretical physics. 
This is related to their role in description of interplay between representations of 
linear and symmetric groups on one side  
and (super)integrability properties of the non-perturbative partition functions
on the other.

Schur functions $S_R\{p\}$ depend on infinite set of time-variables $p_k$, $k=1,\ldots,\infty$
and are labeled by Young diagrams $R$ -- the ordered integer partitions of their size $|R|$,
$R=r_1\geq r_2 \geq \ldots \geq r_{l(R)}>0$, $|R|=r_1+r_2+\ldots + r_{l(R)}$.

\bigskip

{\bf 2.} In this paper we pose a very simple question.
The number of Young diagrams, 
i.e. of partitions or of irreducible representations 
of linear $sl(\infty)$ and symmetric $S(\infty)$ groups, 
have a very simple generating function
\be 
Z(q):=\sum_{R} q^{|R|} = \prod_{n=1} \frac{1}{1-q^n}
\ee
It is natural to ask, if this sum can be somehow related to Cauchy identity \cite{Cauchy}
\be 
\sum_R q^{|R|}\underbrace{S_R\{p\}S_R\{\bar p\}}_{1\ ?} = \exp\left(\sum_k q^k\frac{ p_k\bar p_k}{k}\right)
\ee
which has the same summation domain, but the weight includes a product of two Schur functions. 
The questions is, can we make this weight unity?

At the level $|R|=1$ this looks simple:  $S_{[1]}=p_1$ and it is enough to put $p_1=1$.
The difficulty becomes obvious already at the level $|R|=2$, where there are just two 
diagrams and two Schur functions
\be 
S_{[2]} = \frac{p_2+p_1^2}{2}, \ \ \ \ S_{[1,1]} = \frac{-p_2+ p_1^2}{2}
\ee
Then $S_{[2]} = S_{[1,1]} = 1$ implies that $p_2=0$ and $p_1=\sqrt{2}$, 
which is different from $p_1=1$.
Thus there is no way to choose times so that all $S_R\{p\}=1$.
In general the coefficient of $p_1$ is a non-trivial combinatorial quantity
$S\{\delta_{k,1}\}=d_R$, which for symmetric representations is $d_{[r]} = \frac{1}{r!}$,
and all $p_1=\sqrt{r!}$ with different $r$ are different.

\bigskip

{\bf 3.}
However, in quantum field theory there is a simple way out:
instead of evaluating $S_R$ at given values of arguments $p_k$
one can consider {\it averages}.
The question then is if we can find a measure $d\mu\{p\}$, such that
\be 
\left< S_R \right> := \int S_R\{p\} d\mu\{p\} = 1 \ \ \ \ \forall R
\label{unitave}
\ee
In our above example we get
\be 
\left<p_1\right> = 1,  \ \ \ \ \left<p_1^2\right>=2, \ \ \ \  \left<p_2\right> = 0
\ee

\bigskip

{\bf 4.}  It is straightforward to consider further examples at other levels $|R|$, 
what leads to  the following conclusions about the relevant measure:

\begin{itemize}

\item
The variables $p_k$ are all independent:
\be 
\left<\prod_k p_k^{n_k} \right> = \prod_k \left< p_k^{n_k}\right>
\ee
i.e. the measure is a product $\boxed{d\mu\{p\} = \prod_k d\mu_k(p_k)}$

\item 
The first averages, which are implied by (\ref{unitave}), are
\be 
\begin{array}{c|cccccccccccc}
n&  0&1&2&3&4&5&6&7&8&9&10& \ldots \\ \hline
\left<p_1^n\right> & 1&{\bf 1}&{\bf 2}&{\bf 4}&{\bf 10}&{\bf 26}&76&232&764&2620&9496& \\ \hline
\left<p_2^n\right> & 1&0&2&0&12&0&120&0&1680&0&30240& \\ \hline
\left<p_3^n\right> & 1&1&4&10&46&166&856&3844&21820&114076&703216& \\ \hline
\left<p_4^n\right> & 1&0&4&0&48&0&960&0&26880&0&967680& \\ \hline
\left<p_5^n\right> & 1&1&6&16&106&426&3076&15856&123516&757756&6315976& \\ \hline
\left<p_6^n\right> & 1&0&6&0&108&0&3240&0&136080&0&7348320& \\ \hline
\left<p_7^n\right> & 1&1&8&22&190&806&7456&41308&406652&2719900&28338976& \\ \hline
\ldots
\end{array}
\ee
We note in passing that the boldfaced sub-sequence is dear to the heart of every string physicist.
Also amusing is that the entire first line $\left<p_1^n\right> $ 
appears consisting of the numbers of Young {\it tableaux} -- 
this is the  sequence   A000085  from ref.\!\cite{seqs},
where a number of alternative interpretations is also listed.
There is no obvious reason for  these remarkable properties in the given context.

\item
Looking at this table, one can observe the simple recursions
\be 
\left<p_{2m-1}^n\right> = \left<p_{2m-1}^{n-1}\right> + (2m-1)(n-1)\left<p_{2m-1}^{n-2}\right>
\label{oddave}
\ee
for odd times and 
\be 
\left<p_{2m}^{2n}\right> = 2m(2n-1) \left<p_{2m}^{2n-2}\right>
= (2m)^n(2n-1)!!
\ee
for even times. 
The one for $m=1$ is a well known identity for the numbers of the Young {\it tableuax}.
 
\item
Eq.(\ref{oddave}) is reminiscent of the three-term relation for orthogonal polynomials
\cite{orthopols,UFN3}. 
Moreover, from 
\be 
\left<(p-1)p^{n-1}\right>= \left<p^{n}+p^{n-1}\right> 
= (2m-1)(2n-1)\left<p^{n-2}\right> = (2m-1)\left<\frac{\p}{\p p} p^{n-1}\right>
\ee
and using integration by parts,
one deduces  the differential equation for the measure
\be 
(2m-1)\frac{\p du_{2m-1}(p)}{\p p} = -(p-1) d\mu_{2m-1}(p)  \ \Longrightarrow \ 
\nn \\ 
\boxed{
d\mu_{2m-1}(p)  = \frac{1}{\sqrt{2(2m-1)\pi}} \exp\left(-\frac{(p-1)^2}{2(2m-1)}\right)dp
} 
\ee
i.e. the measure is Gaussian in the time-variable(!). 
Similarly, 
\be 
\boxed{
d\mu_{2m}(p) =\frac{1}{2\sqrt{\pi m}} \exp\left(-\frac{p^2}{4m}\right)dp 
}
\ee

\end{itemize}

\bigskip

{\bf 5.} Coming back to original question. we obtain the identity
\be 
Z(q)=\prod_k \frac{1}{1-q^k} = \sum_R q^{|R|} =
\sum_R q^{|R|}\underbrace{\left<S_R\{p\}\right>}_{1}\underbrace{\left<S_R\{\bar p\}\right>}_{1}
= \left< \exp\left(\frac{q^kp_k\bar p_k}{k}\right)\right>
\ee
in the form
\vspace{-0.3cm}
\be 
\boxed{
\prod_{k=1}^\infty(1-q^k) \underbrace{ \int \int e^{\frac{q^k p\bar p}{k}} d\mu_k\{p\}d\mu_k\{\bar p\}}_{Z_k(q)}
=1 
}
\ee
One could hope that this is a termwise identity, but actually it is not:
each individual $(1-q^k)\cdot Z_k(q)\neq 1$, only the total product is unity.
For example,
\be 
\begin{array}{ccccccccccccc}
Z_1(q) =& 1 &\!\!+ q + 2q^2 &+ \frac{8}{3}q^3& + \frac{25}{6}q^4& + \frac{169}{30}q^5 
&+ \frac{361}{45}q^6 &+ \frac{3364}{315}q^7 &+ \frac{36481}{2520}q^8 
&+& \ldots 
\nn \\  \nn \\
Z_2(q) =& 1& &&+ \frac{1}{2} q^4 &&&&+ \frac{3}{8}q^8& + &\ldots 
\nn \\ 
Z_3(q) =& 1 &&+ \frac{1}{3} q^3 &&& + \frac{8}{9}q^6 &&&  +& \ldots
\nn\\ 
Z_4(q) =& 1& &&&&&&+ \frac{1}{2} q^8 & +& \ldots 
%\frac{3}{8}q^16 + \ldots 
\nn \\
Z_5(q) =& 1& &&&+ \frac{1}{5}q^5 &&&& +& \ldots 
\nn \\
Z_6(q)=&1& &&&&&&& +& \ldots
\nn \\
Z_7(q)=&1& &&&&& + \frac{1}{7}q^7 &&+&\ldots 
\nn \\
Z_8(q) =& 1& &&&&&&& +& \ldots 
\nn \\
\ldots
\nn \\ \hline \nn \\
\prod_{k=1}\frac{1}{1-q^k}=&1& \!\! +q+2q^2 & + 3q^3 & +5q^4 & + 7q^5 & + 11 q^6 & +15q^7 & +22 q^8 &  +&\ldots
\end{array}
\ee
Note that the first lines should be {\it multiplied}, not added, in order to coincide with the last line.

\bigskip

{\bf 6.}
This note reports a funny observation about eliminating the factors $d_R=S_R\{\delta_{k,1}\}$ 
from Cauchy identity and efficiently averaging Schur functions to unities --
all at once.
Somewhat surprisingly, Gaussian measures appear sufficient to solve this problem. 
It can possess interesting generalizations to Jack and Macdonald polynomials \cite{Mac}
and, most interesting to the still hypothetical $3$-Schur functions \cite{3Schurs},
associated to plain partitions, when Cauchy identity would get related to the
MacMahon generating function $\prod_n  (1-q^k)^{-k}$.
%$\prod_k\frac{1}{(1-q^k)^k}$
As a bonus we get an integral formula for the number of Young {\it tableaux} --
they appear to be simple Gaussian averages.
Intriguing are the possible relations to matrix models \cite{UFN3}
and to their superintegrability properties \cite{si} --
but this requires restriction to finite matrix size $N$
(what nullifies some of the Schur functions)
and a matrix-model interpretation of our new measures, 
which are Gaussian in times rather than in matrix and eigenvalue variables $X={\rm diag}(x_i)$,
usually used in matrix models {\it a la} \cite{si}  to parameterize the Miwa locus $p_k=\tr X^k$.

\section*{Acknowledgements}

This work is  supported by the Russian Science Foundation (Grant No.21-12-00400).

\end{document}